\documentclass[preprint,amsmath,amssymb,superscriptaddress,aps,prx,bibnotes]{revtex4-1}
\usepackage[dvipdfmx]{graphicx}
\usepackage{booktabs}
\usepackage{comment}
\usepackage{sidecap}
\usepackage{dcolumn}
\usepackage{bm}
\usepackage{here}
\usepackage{bm}
\usepackage[dvipdfmx]{hyperref}
\hypersetup{
 setpagesize = false,
 colorlinks = true,
 citecolor = blue,
 linkcolor = blue,
 urlcolor = blue,
 }
\usepackage[all] {hypcap}

\begin{document}

\preprint{APS/123-QED}

\title{Electronic charge transfer driven by spin cycloidal structure}

\author{Y. Ishii}
\email{yuta.ishii.c2@tohoku.ac.jp}
\altaffiliation{Present address: Department of Physics, Tohoku University, Sendai, Miyagi 980-8578, Japan.}
\affiliation{Institute of Materials Structure Science (IMSS), High Energy Accelerator Research Organization (KEK), Tsukuba, Ibaraki 305-0801, Japan}
\author{S. Horio}
\affiliation{Institute of Multidisciplinary Research for Advanced Materials, Tohoku University, Aoba Sendai 980-8577, Japan}
\author{Y. Noda}
\affiliation{Institute of Multidisciplinary Research for Advanced Materials, Tohoku University, Aoba Sendai 980-8577, Japan}
\author{M. Hiraishi}
\affiliation{Institute of Materials Structure Science (IMSS), High Energy Accelerator Research Organization (KEK), Tsukuba, Ibaraki 305-0801, Japan}
\author{H. Okabe}
\affiliation{Institute of Materials Structure Science (IMSS), High Energy Accelerator Research Organization (KEK), Tsukuba, Ibaraki 305-0801, Japan}
\author{M. Miyazaki}
\affiliation{Graduate School of Engineering, Muroran Institute of Technology, Muroran, Hokkaido 050-8585, Japan}
\author{S. Takeshita}
\affiliation{Institute of Materials Structure Science (IMSS), High Energy Accelerator Research Organization (KEK), Tsukuba, Ibaraki 305-0801, Japan}
\author{A. Koda}
\affiliation{Institute of Materials Structure Science (IMSS), High Energy Accelerator Research Organization (KEK), Tsukuba, Ibaraki 305-0801, Japan}
\author{K. M. Kojima}
\affiliation{TRIUMF Centre for Molecular and Material Science (TRIUMF-CMMS) and Quantum Matter Institute, University of British Columbia, Vancouver, BC  V6T 2A3, Canada}
\author{R. Kadono}
\affiliation{Institute of Materials Structure Science (IMSS), High Energy Accelerator Research Organization (KEK), Tsukuba, Ibaraki 305-0801, Japan}
\author{H. Sagayama}
\affiliation{Institute of Materials Structure Science (IMSS), High Energy Accelerator Research Organization (KEK), Tsukuba, Ibaraki 305-0801, Japan}
\author{H. Nakao}
\affiliation{Institute of Materials Structure Science (IMSS), High Energy Accelerator Research Organization (KEK), Tsukuba, Ibaraki 305-0801, Japan}
\author{Y. Murakami}
\affiliation{Institute of Materials Structure Science (IMSS), High Energy Accelerator Research Organization (KEK), Tsukuba, Ibaraki 305-0801, Japan}
\author{H. Kimura}
\affiliation{Institute of Multidisciplinary Research for Advanced Materials, Tohoku University, Aoba Sendai 980-8577, Japan}

\begin{abstract}

Muon spin rotation and resonant soft X-ray scattering experiments on 
prototype multiferroics $R$Mn$_2$O$_5$ ($R$ = Y, Sm) are used to demonstrate that the local electric displacements are 
driven by the spin-current (SC) mechanism.
Small local electric displacements were evaluated by 
observing spin polarization at ligand O ions, 
for which implanted muons served as an extremely sensitive probe. 
Our results for YMn$_2$O$_5$ provide evidence that 
the spin polarization of O ions forming a spin cycloid chain with Mn spins increases in
proportion to the vector spin chirality (${\bf S}_i \times {\bf S}_j$) of the Mn ions.
This relationship strongly indicates that the charge transfer between O and Mn ions is 
driven by the SC mechanism, which leads to the ferroelectricity accompanying O spin polarization.

\end{abstract}

\pacs{Valid PACS appear here}

\maketitle

\section{\label{sec:level1} Introduction}

Manipulating the electronic state of materials by tuning their 
magnetic state is key to the development of new devices with multifunctional properties.
To this end, multiferroic materials with coupling between magnetism and 
ferroelectricity provide promising playing grounds \cite{Tokura}.
A number of studies on these multiferroics 
have prompted the development of theoretical models for 
the microscopic mechanism of magnetically driven ferroelectricity, such as 
the "spin-current" (SC) and "exchange-striction" (ES) models,
in which a cycloidal spin chain and a collinear spin alignment, 
respectively, break the inversion symmetry 
\cite{HKatsura2005,YYamasaki2006, BLorenz2007}.

Elucidating these local displacements associated with 
the electric polarization is critical for examining these models 
and gaining better understanding of multiferroics. 
However, previous attempts have achieved limited success 
because of the experimental difficulties associated with
observing these local deviations, which are expected to be small in proportion to 
the electric polarization in most multiferroic materials.
Recently, some authors have suggested that the charge transfer between 
transition-metal cations and oxygen ligands 
would represent the primary contribution to the ferroelectricity 
and that the charge transfer would also accompany oxygen spin polarization
\cite{ASMoskvin2008_1,ASMoskvin2008_2}.
Hence, resonant soft X-ray scattering (RSXS) experiments 
at the O $K$-edge were performed to observe the magnetic order 
of O ions for prototype multiferroic materials 
$R$Mn$_2$O$_5$ and $R$MnO$_3$ ($R$ = rare earth)
\cite{TAWBeale2010,SPartzsch2011,RAdeSouza2011,SWHuang2016,YIshii2018,YIshii2019}.
In particular, charge transfer was inferred to contribute to ferroelectricity 
in YMn$_2$O$_5$ and 
Tb$_{0.5}$Gd$_{0.5}$Mn$_2$O$_5$ 
on the basis of a one-to-one correspondence in the 
temperature and external magnetic field dependence, respectively, 
between the electric polarization and the magnetic scattering intensity of the O ions
\cite{SPartzsch2011,YIshii2019}.
Despite these efforts, RSXS measurements preclude quantitative discussion
of the O spin polarization in evaluating the contribution of 
charge transfer to the ferroelectricity.

In the present  paper, 
we report the extraction of information on the spin polarization of ligand 
O ions via a synergetic use of RSXS and muon spin rotation ($\mu$SR)
as mutually complementarily techniques,
as demonstrated for YMn$_2$O$_5$.
While the magnetic order associated with O spin polarization is 
inferred from the magnetic reflection of O 2$p$ electrons via 
the O $K$ edge RSXS,
the magnitude of O spin polarization is estimated from the local magnetic field 
measured by $\mu$SR; positive muons, which are regarded as pseudo-hydrogen atoms, exhibit a strong 
tendency to form OH bonds with oxygen ligands in nonmetallic oxides, 
thus serving as a sensitive probe for spin polarization.

We use SmMn$_2$O$_5$ as a reference for comparison to 
YMn$_2$O$_5$ because they exhibit contrasting 
magnetic and ferroelectric properties as well as contrasting
oxygen spin states
\cite{IKagomiya2001,YNoda2008,YIshii2016}.
Both of these materials undergo a sequence of dielectric and magnetic phase transitions.
In the case of YMn$_2$O$_5$, 
an incommensurate antiferromagnetic phase 
appears below 45 K. 
Electric polarization occurs upon cooling to 40 K, 
concomitant with a commensurate magnetic order (CM phase)
with a magnetic propagation vector ${\bm q}_{\rm M}$ = (1/2, 0, 1/4).
With a further decrease in temperature to less than 18 K, the electric polarization rotates in the opposite direction, which is related to the transition to a low-temperature
incommensurate magnetic (LT-ICM) phase.
The magnetic structure has been clarified in previously reported neutron scattering
experiments, where an approximately collinear magnetic alignment in the 
$ab$ plane and spin cycloidal chain of Mn$^{4+}$ moments along the 
$c$-axis were realized \cite{HKimura2007,JHKim2008}.  
This magnetic structure leads to the scenario that the ES model is responsible for the
ferroelectricity in the CM phase, whereas the SC model is in effect for the 
LT-ICM phase \cite{SWakimoto}.
By contrast, an incommensurate magnetic (ICM) phase appears below 34 K 
in SmMn$_2$O$_5$.
The electric polarization is enhanced upon the onset of a CM phase
below 26 K with a perfect collinear magnetic structure 
with ${\bm q}_{\rm M}$ = (1/2, 0, 0) \cite{YIshii2016,GYahia2017}.
This magnetic structure suggests that the electric polarization in the CM phase
is attributable to the ES model.

\section{\label{sec:level2} Experiment details}

Single crystals of YMn$_2$O$_5$ and SmMn$_2$O$_5$ were grown by the PbO--PbF$_2$ flux method
\cite{BMWankly1972}.
We performed RSXS experiments in which measurements at the O $K$-edge ($E \sim$ 530 eV) were carried out at the BL-16A and BL-19B beamlines
\cite{HNakao2014} at the Photon Factory, KEK, Japan.  
Conventional $\mu$SR experiments were performed at the M15 and M20 beamlines at TRIUMF, Canada.
Zero-field (ZF) and transverse-field (TF)
$\mu$SR were conducted using the LAMPF spectrometer and NuTime spectrometer.


\section{\label{sec:level3} Results and discussion}
\subsection{\label{sec:level1} RSXS measurements}

\begin{figure}[t]
\centering
\includegraphics [clip,width=6cm]{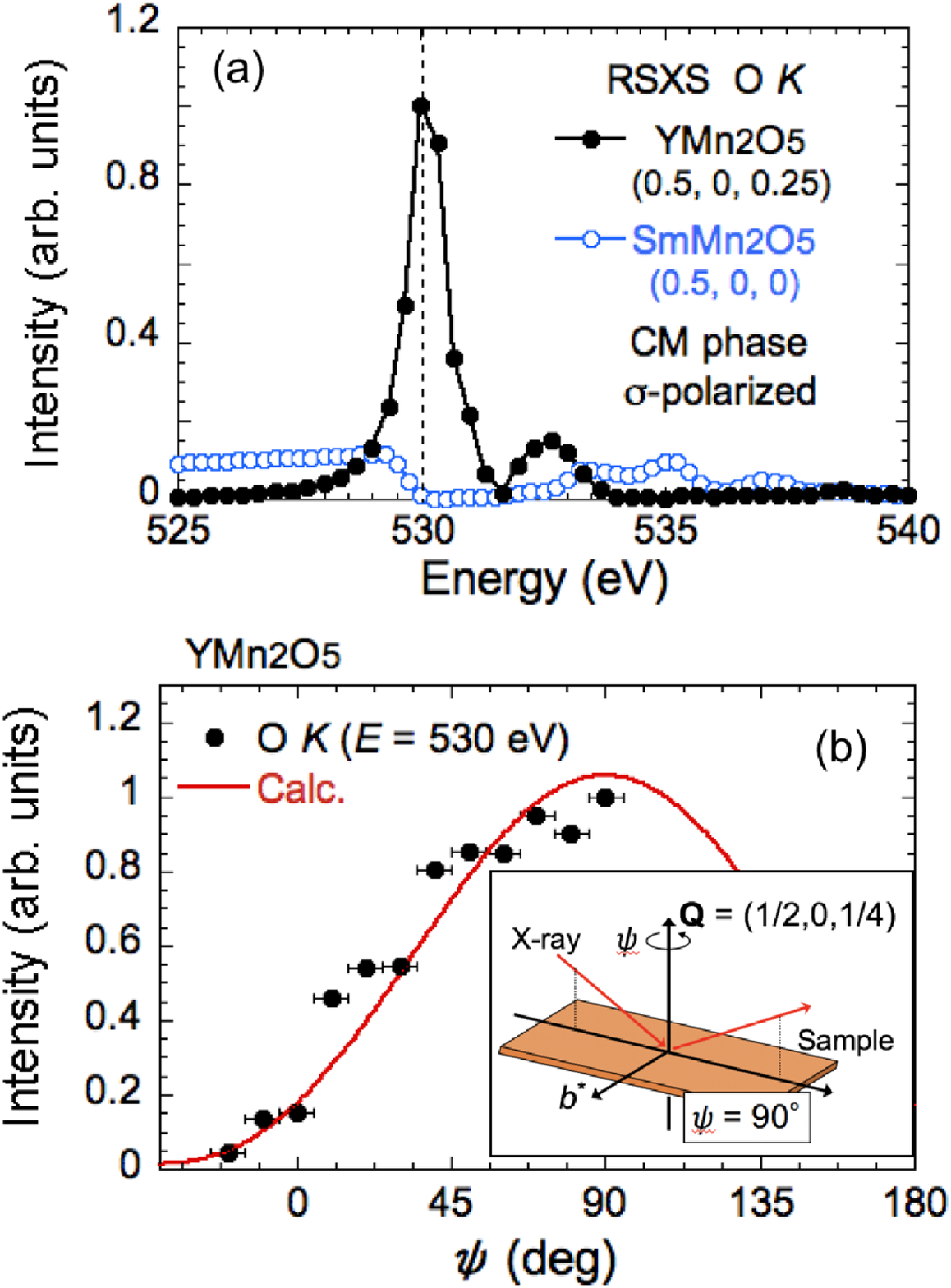}
\caption{(Color online). 
(a) Energy spectra of RSXS around the O $K$ edge in $R$Mn$_2$O$_5$ ($R$ = Y, Sm).
Each measurement was performed in the CM phase (26 K for YMn$_2$O$_5$, 18 K for SmMn$_2$O$_5$) of each material.
(b) Azimuth angle dependence of RSXS intensity at O $K$-edge ($E = 530$ eV) in YMn$_2$O$_5$.
The red curve represents the calculated values with the magnetic structure of O2 ions (shown in Fig.\ \hyperref[fig:YMO_mSR]{\ref*{fig:YMO_mSR} (c)}).
The inset shows the scattering geometry at azimuthal angle $\psi = 90^{\circ}$.
By definition, $ \psi$ = 90$^{\circ}$ indicates that the $b^{*}$-axis is perpendicular to the scattering plane.
}
\label{fig:SmY_Escan_azimuth} 
\end{figure}

RSXS energy spectra around the O $K$-edge obtained 
for the CM phase of YMn$_2$O$_5$
and SmMn$_2$O$_5$ are shown in Fig.\ \hyperref[fig:SmY_Escan_azimuth]{\ref*{fig:SmY_Escan_azimuth} (a)},
which reproduce those reported previously \cite{SPartzsch2011,RAdeSouza2011,YIshii2018}.
The incident X-rays were polarized perpendicular to the scattering plane 
($\sigma$-polarized).
In the spectrum of YMn$_2$O$_5$, a well-defined peak is observed at $E \approx$ 530 eV,
which indicates the presence of spin polarization of O sites via charge transfer from 
Mn to O ions. 
By contrast, no such peak is observed in the spectra for SmMn$_2$O$_5$,
strongly suggesting the absence of the O spin polarization via Mn--O hybridization.
Thus, we confirmed the oxygen ligands in the CM phase of YMn$_2$O$_5$ and SmMn$_2$O$_5$ have different electronic states. This conclusion is further supported by the azimuthal angle dependence of the intensity 
at $E$ = 530 eV measured for YMn$_2$O$_5$ [see Fig.\ \hyperref[fig:SmY_Escan_azimuth]{\ref*{fig:SmY_Escan_azimuth} (b)}].
The resonant intensity increases as the azimuthal angle 
approaches $\psi$ = 90$^{\circ}$.
This observed angle dependence is in reasonable agreement with that calculated 
for the magnetic structure of O ions suggested by the present $\mu$SR experiment (see below).

\subsection{\label{sec:level2} Muon stopping sites}

\begin{figure}[t]
\centering
\includegraphics [clip,width=8.5cm]{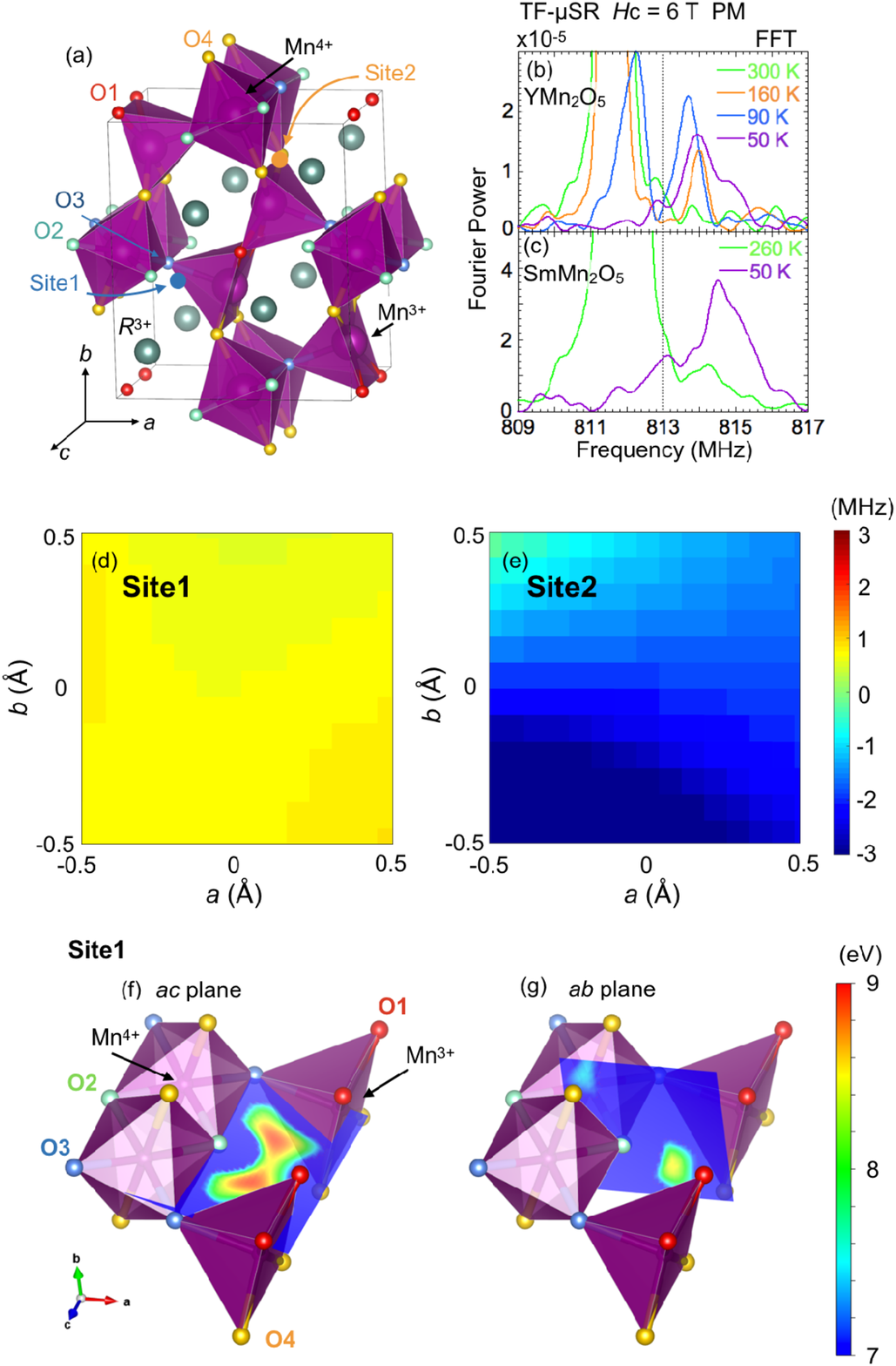}
\caption{(Color online) Fig.~S1: (a) Crystal structure of $R$Mn$_2$O$_5$. 
Site1 and Site2  represent candidate muon stopping sites indicated by VASP calculations.
(b)(c) FFT spectra of TF-$\mu$SR at each temperature under an external magnetic field along the $c$-axis for YMn$_2$O$_5$ and SmMn$_2$O$_5$. The dashed line represents 813 MHz.  (d)(e) Contour map of the local magnetic field (shown in terms of the frequency shift $\gamma_\mu{\bm B}_{\rm loc}/2\pi$) in the $ab$ plane around Site1 and Site2.
Positive (negative) indicates parallel (antiparallel) to the external magnetic field.
(f)(g) Hartree potential in the $ac$ and $ab$ planes around Site1. 
}
\label{fig:StTFmVasp} 
\end{figure}
For the $\mu$SR measurement, we first narrowed down candidate sites 
for the interstitial muons in the $R$Mn$_2$O$_5$ structure 
by investigating the Hartree potential obtained by first-principle calculations
based on density functional theory (DFT).
Fig.\ \hyperref[fig:StTFmVasp]{\ref*{fig:StTFmVasp} (a)} shows 
the crystal structure containing a unit cell in the paramagnetic phase of $R$Mn$_2$O$_5$.
This structure has two independent sites for Mn (Mn$^{4+}$ and Mn$^{3+}$)
and four independent sites for O (sites O1 through O4).
They form edge-sharing Mn$^{4+}$--O$_6$ octahedra running along the $c$-axis,
whereas a pair of Mn$^{3+}$--O$_5$ pyramids links the Mn$^{4+}$--O$_6$ chains in the 
$ab$ plane. 
The Hartree potential for this structure was calculated using 
the Vienna Ab initio Simulation Package (VASP) code \cite{Vasp}
to explore the potential minima for a positive charge introduced by muons.
Site1 and Site2 in Fig.\ \hyperref[fig:StTFmVasp]{\ref*{fig:StTFmVasp} (a)}
correspond to the obtained local potential minima located near O2 and O4 sites,
respectively.

These candidate sites were subsequently examined on the basis of the muon Knight shift data obtained 
by TF-$\mu$SR measurements under an external magnetic field ($H_c$ = 6 T)
applied along the $c$-axis.
The frequency shift in the paramagnetic (PM) phase is 
determined by the local magnetic field (${\bm B}_{\rm loc}$)
at the muon site, which is given by the vector sum of the dipolar fields 
generated by local magnetic moments:
\begin{equation}
{\bm B}_{\rm loc} =\sum_jA^{\alpha\beta}_j\mu^\beta_j,\label{bloc0}
\end{equation}
 where the summation runs through the $j$-th Mn magnetic moment ${\bm \mu}_j=(\mu_j^x,\mu_j^y,\mu_j^z)$ located at ${\bm r}_j=(x_j,y_j,z_j)$ from a given muon site, $A^{\alpha\beta}_j$ is the corresponding dipole tensor (whose $z$-axis is chosen parallel with the crystalline $c$-axis), which is expressed by the equation
\begin{equation}
A^{\alpha\beta}_j=\frac1{r_j^3}\left(\frac{3\alpha_j\beta_j}{r_j^2}-\delta_{\alpha\beta}\right)\quad(\alpha, \beta=x,y,z).
\label{dipten}
\end{equation}
Assuming that the Mn moments are aligned parallel to $H_c$ in the PM phase, then
\begin{equation}
{\bm B}_{\rm loc} =\frac{\chi_c}{N_A\mu_B}\sum_jA^{zz}_j,
\label{bloc}
\end{equation}
where $\chi_c$ is the bulk magnetic susceptibility, $N_A$ is Avogadro's number, and $\mu_B$ is the Bohr magneton.

Figures \hyperref[fig:StTFmVasp]{\ref*{fig:StTFmVasp} (b)(c)} show 
fast Fourier transforms (FFTs) of the TF-$\mu$SR spectra
 in the PM phase ($T > 50$ K) of $R$Mn$_2$O$_5$, where
the dashed line represents the null shift ($\gamma_\mu H_c/2\pi=813$ MHz, where $\gamma_\mu/2\pi=135.53$ MHz is the muon gyromagnetic ratio).
A strong peak was observed around 811.5 MHz at ambient temperature for both compounds.
With decreasing temperature below $\sim$160 K, an additional well-defined peak emerges at approximately 814 MHz,
whereas the peak at 811.5 MHz diminishes and then disappears at $\sim$50 K. 
It is inferred from these observations that there are two different muon stopping sites in the unit cell,
where the ${\bm B}_{\rm loc}$ probed by muons is parallel or antiparallel  to the direction of the external magnetic field, respectively.
The intensity of Fourier power at 50 K decreases compared with that at ambient temperature, 
likely because of a strong relaxation near the N$\acute{\rm e}$el temperature. 

These results were compared with the ${\bm B}_{\rm loc}$ calculated using Eq.~(\ref{bloc}) \cite{KMKojima2004} for Site1 and Site2 in YMn$_2$O$_5$. 
Figures \hyperref[fig:StTFmVasp]{\ref*{fig:StTFmVasp} (d)(e)}
show the contour map of ${\bm B}_{\rm loc}$ in the $ab$ plane around Site1 and Site2,
where positive (negative) values represent ${\bm B}_{\rm loc}$ parallel (antiparallel) 
to the direction of an external field. 
Both the magnitude and sign of ${\bm B}_{\rm loc}$ for Site1 and Site2 are in good agreement  with the experimental observations; thus, these two sites can be reasonably assigned as the most likely muon sites. 
This result further suggests that the muon occupancy is shifted  
from Site2 to Site1 as temperature decreases in the PM phase.
A recent X-ray absorption spectroscopy (XAS) study of TbMn$_2$O$_5$ 
implies the occurrence of local buckling at the O4 site 
below $T \sim 180$ K in the PM phase, as evidenced by the thermal evolution of 
the pair-distribution function for a Tb--O bond \cite{TATyson2007}.
The local buckling at the O4 site likely occurs also in YMn$_2$O$_5$, where it causes a change of the local electrostatic potential around Site2, leading to a shift of muon occupancy from Site2 to Site1 at temperatures below $T \sim 160$ K.
Figures \hyperref[fig:StTFmVasp]{\ref*{fig:StTFmVasp} (f)(g)}
show the shape of the Hartree potential 
around Site1, as obtained by VASP calculations; this result
suggests that 
the candidate area for the muon stopping site is broadened in the $ac$ plane.
Thus, we conducted ZF-$\mu$SR for SmMn$_2$O$_5$ to further narrow the candidate muon 
sites.

\subsection{\label{sec:level2} ZF-$\mu$SR measurements for SmMn$_2$O$_5$}
\begin{figure*}[t!]
\centering
\includegraphics [clip,width=18cm]{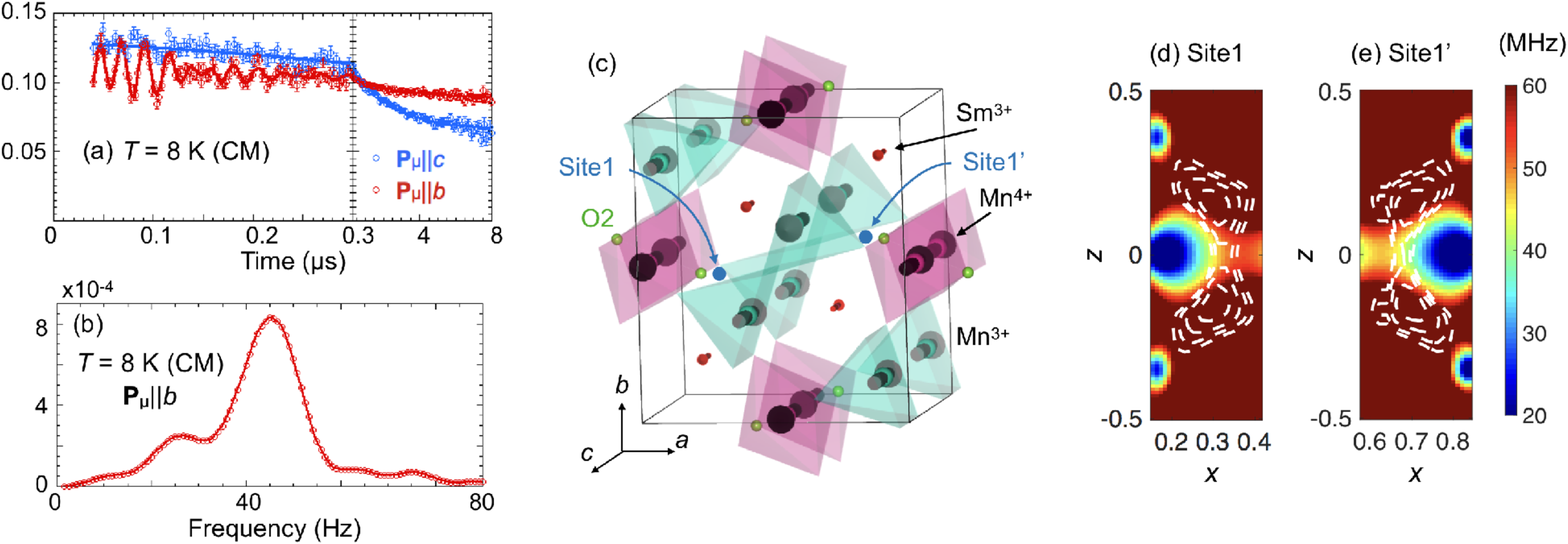}
\caption{(Color online) 
(a)  Time spectra of ZF-$\mu$SR for SmMn$_2$O$_5$ at 8 K in the CM phase, 
and (b) FFT of the time spectrum with  ${\bm P}_{\mu} || b$.
(c) Magnetic structure in the CM phase of SmMn$_2$O$_5$.
Red and black arrows represent the magnetic moments of Sm and Mn ions, respectively.
Site1$^{(')}$ correspond to muon stopping sites.
(d)(e) Contour map of ${\bm B}_{\rm loc}$ in the $ac$ plane around Site1 and Site1$^{'}$. 
The magnitude is represented by the corresponding muon spin precession frequency 
($\gamma_\mu B_{\rm loc}/2\pi$).
Dashed lines represent the contour of the Hartree potential 
shown in Fig.\ \hyperref[fig:StTFmVasp]{\ref*{fig:StTFmVasp} (f)}.
}
\label{fig:SMO_mSR} 
\end{figure*}
Figure \hyperref[fig:SMO_mSR]{\ref*{fig:SMO_mSR} (a)}
shows time spectra at 8 K (CM phase)
with the initial muon spin polarization [${\bm P}_{\mu}(t = 0)$]
parallel to the $c$- and $b$-axes.
The muon spin precession signal is clearly observed with ${\bm P}_{\mu} || b$.
The corresponding FFT spectrum in Fig.\ \hyperref[fig:SMO_mSR]{\ref*{fig:SMO_mSR} (b)}
suggests two frequency components. 
The spectrum was subsequently analyzed by curve fitting \cite{Fitting}; the deduced frequencies are shown in Table \hyperref[tb:SMO_calc_site1]{\ref*{tb:SMO_calc_site1}} as $B^{\rm exp}$.
In sharp contrast, no precession is observed in the spectrum with ${\bm P}_{\mu} || c$ within the experimental error. 
In addition, the curve fitting analysis failed to reproduce the spectrum. 
Because muon spin precession was induced by ${\bm B}_{\rm loc}\perp {\bm P}_{\mu}$,
these observations indicate that 
${\bm B}_{\rm loc}$ at the relevant muon site is parallel to the $c$-axis. 
We calculated ${\bm B}_{\rm loc}$ 
for the magnetic structure of Sm and Mn ions 
shown in Fig.\ \hyperref[fig:SMO_mSR]{\ref*{fig:SMO_mSR} (c)}
reported in the literature \cite{YIshii2016,GYahia2017}.
In this magnetic structure, 
Site1 divides into two inequivalent sites, which we refer to
as Site1 and Site1$^{'}$. 
The calculated values of ${\bm B}_{\rm loc}$ around Site1 and Site1$^{'}$
are shown in Figs.\ \hyperref[fig:SMO_mSR]{\ref*{fig:SMO_mSR} (d)(e)}.
Dashed lines in these figures represent the contour of the Hartree potential 
shown in Fig.\ \hyperref[fig:StTFmVasp]{\ref*{fig:StTFmVasp} (f)}.
In the area indicated by the VASP calculation, 
the sites where the calculated ${\bm B}_{\rm loc}$ are in good agreement with
the experimental values are determined  to be 
$(0.301(1), 0.401(2), 0.00(1))$ and $(0.718(1), 0.599(2), 0.00(1))$, respectively.
The calculated ${\bm B}_{\rm loc}$ at these sites are also summarized in Table 
\hyperref[tb:SMO_calc_site1]{\ref*{tb:SMO_calc_site1}} as $B^{\rm calc}$,
which are in agreement with the experimental values.
Here, we stress that $B^{\rm calc}$ reproduces $B^{\rm exp}$
 without the assumption of spin polarization at the O sites,
 which is consistent with the results of the RSXS experiments
\cite{YIshii2018}.

\begin{table}[t!]
\centering
\caption{Magnitude and direction of the local magnetic field ($B_{\rm loc}$) at Site1$^{(')}$ in the CM phase
in SmMn$_2$O$_5$, where $B^{\rm exp}$ is deduced from ZF-$\mu$SR experiments and $B^{\rm calc}$ are calculated values. 
The magnitude is represented by the
corresponding muon spin precession frequency ($\gamma_\mu B_{\rm loc}/2\pi$).}
{\tabcolsep = 3mm
 \begin{tabular}{c c c c c c}  \\ \toprule
Site& $B^{\rm exp}$  (MHz)  & $B^{\rm exp}_{\rm dir}$  & $B^{\rm calc}$ (MHz) & $B^{\rm calc}_{\rm dir}$ \\ \midrule 
1 &44.4(2)& $c$-axis &44.5(2) & $c$-axis\\
1$^{'}$ &34.2(2)&  $c$-axis &34.2(2) & $c$-axis\\ \bottomrule
\end{tabular} }
\label{tb:SMO_calc_site1}  
 \end{table} 

\begin{figure*}[t!]
\centering
\includegraphics [clip,width=14cm]{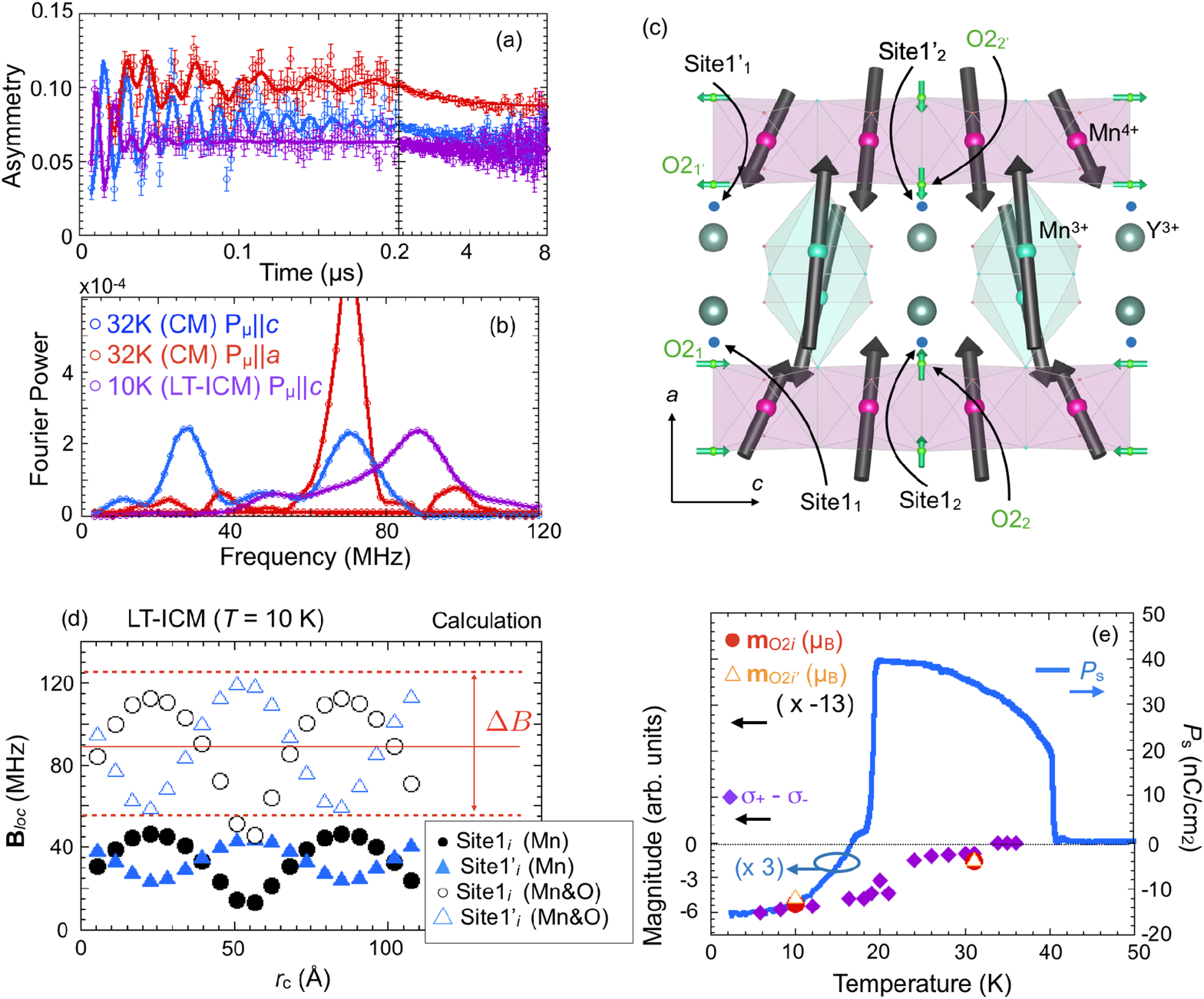}
\caption{(Color online). (a) Time spectra and (b) FFT spectra of ZF-$\mu$SR for YMn$_2$O$_5$ at 32 K and 10 K.
(c) Magnetic structure of Mn and O2 ions in the CM phase of YMn$_2$O$_5$ projected onto the $ac$ plane. 
Black and green arrows represent the magnetic moments of Mn and O2 ions, respectively. 
The magnitude of magnetic moments at O2 sites is exaggerated for clarity.
Filled blue circles represent 
muon stopping sites Site1$^{(')}_{i}$ ($i = 1, 2$).
(d) Calculation of ${\bm B}_{\rm loc}$ 
at Site1$^{(')}_{i}$ for $T$ = 10 K. 
Filled circles (triangles) 
represent calculated values at Site1$^{(')}_{i}$ with only Mn magnetic moments.
Open circles and triangles
represent calculated values with Mn and O magnetic moments. 
The data are plotted over different chemical unit cells along the $c$-axis. 
The line represents the magnitude of 
${\bm B}_{\rm loc}$ ($\sim$ 88.9 MHz), 
and dashed lines represent
the distribution width  $\Delta B \sim$ 70 MHz.
 (e) Temperature dependence of the amplitude of 
 the oxygen magnetic moments at O2$_i$ and O2$_{i^{'}}$ sites (${\bf m}_{\rm O2_{\it i}}$ and ${\bf m}_{\rm O2_{{\it i}^{'}}}$), $\sigma_{+}-\sigma_{-}$, and the electric polarization ($P$s). The $\sigma_{+}-\sigma_{-}$ data are taken from Ref.\ \cite{SWakimoto}.
The values of $P$s corresponding to temperatures below 18 K, ${\bf m}_{\rm O2_{\it i}}$, and ${\bf m}_{\rm O2_{{\it i}^{'}}}$ are exaggerated
for clarity.}
\label{fig:YMO_mSR} 
\end{figure*}

\subsection{\label{sec:level3} ZF-$\mu$SR measurements for YMn$_2$O$_5$}

In the case of ZF-$\mu$SR for YMn$_2$O$_5$, 
muon spin precession signals are clearly observed 
at 32 K (in the CM phase)
with ${\bm P}_{\mu} || c$ and ${\bm P}_{\mu} || a$,
as shown in Fig.~\hyperref[fig:YMO_mSR]{\ref*{fig:YMO_mSR} (a)}.
The FFTs of the spectra shown in 
Fig.\ \hyperref[fig:YMO_mSR]{\ref*{fig:YMO_mSR} (b)} 
show single and two well-defined peaks for ${\bm P}_{\mu} || c$ and ${\bm P}_{\mu} || a$,  respectively. 
The corresponding frequencies obtained by curve fitting ($B^{\rm exp}$) are summarized in Table \hyperref[tb:YMO_calc_site1_Mn]{\ref*{tb:YMO_calc_site1_Mn}}.
Site1 is predicted to divide into four inequivalent sites because of the magnetic symmetry [shown in Fig.\  \hyperref[fig:YMO_mSR]{\ref*{fig:YMO_mSR} (c)}], which we refer to as Site1$^{(')}_{i}$ ($i$ = 1,2).
We first calculated ${\bm B}_{\rm loc}$ at these sites by assuming that the magnetic moments appear only at the Mn sites in the CM phase
obtained by neutron scattering experiments \cite{YNoda}. 
The calculated magnitude and direction of  ${\bm B}_{\rm loc}$ are shown 
 in Table \hyperref[tb:YMO_calc_site1_Mn]{\ref*{tb:YMO_calc_site1_Mn}}
as $B^{\rm calc, Mn}$ and $B^{\rm calc, Mn}_{\rm dir}$, respectively.
These magnitudes differ substantially from the experimental values, though
their directions are consistent.
We next considered the possibility that the magnetic moments appear at the O2 site with their directions 
parallel to a local magnetic field generated by the neighboring Mn moments, 
which would be reasonable because the O spin polarization is induced by the Mn magnetic order.
The presumed magnetic structure for the O2 sites 
is shown in Fig.\ \hyperref[fig:YMO_mSR]{\ref*{fig:YMO_mSR} (c)}.
We found excellent agreement between the calculated values for ${\bm B}_{\rm loc}$ and those deduced experimentally when the O2 moment size was assumed to be $\sim$ 0.1 $\mu_{\rm B}$. These results are summarized in 
 Table \hyperref[tb:YMO_calc_site1_Mn]{\ref*{tb:YMO_calc_site1_Mn}}
as $B^{\rm calc, Mn, O}$ and $B^{\rm calc, Mn,O}_{\rm dir}$.
We also calculated the azimuthal angle dependence 
of resonant intensity at the O $K$-edge for this O2 magnetic structure
by using the resonant magnetic scattering amplitude \cite{JPHill1996}, which is 
 shown as the red curve in Fig.~\hyperref[fig:SmY_Escan_azimuth]{\ref*{fig:SmY_Escan_azimuth} (b)}.
The reasonable agreement with experimental observations
strongly supports the appearance of magnetic moments at the O2 site inferred from the present $\mu$SR measurements.

We also observed muon spin precession signals at 10 K (in the LT-ICM phase)
as shown in Fig.\ \hyperref[fig:YMO_mSR]{\ref*{fig:YMO_mSR} (a)}.
The time spectrum was analyzed by curve fitting \cite{Fitting},
and well reproduces the experimental results
for a single component with 
frequency $\omega/2\pi=88.9(1.0)$ (MHz) and 
relaxation rate $\lambda= 30(10)$ ($\mu{\rm s}^{-1}$).
This large relaxation rate is attributed to a wide distribution of ${\bm B}_{\rm loc}$ resulting from
the incommensurate magnetic structure.
To estimate the distribution width of ${\bm B}_{\rm loc}$,
we assumed that  ${\bm B}_{\rm loc}$ follows an isotropic Gaussian distribution, 
\begin{eqnarray}
f(B_{\rm loc}) \propto \exp(-\frac{|{\bm B}_{\rm loc}|^2}{2\delta^2}),
\label{eq:Gaussdis}  
\end{eqnarray}
where $\delta$ is the dispersion of ${\bm B}_{\rm loc}$.
Provided that the ${\bm B}_{\rm loc}$ is quasi-static in the time scale of $\mu$SR, 
$\delta$ corresponds to the relaxation rate ($\delta \sim \lambda$).
This relationship enables us to estimate the full-width at half-maximum of $f(B_{\rm loc})$ as
$\Delta B = 2\sqrt{2\log{2}} \cdot \lambda \simeq$ 70 MHz,
which is comparable to
the width of the FFT spectrum at 10 K in Fig.~\hyperref[fig:YMO_mSR]{\ref*{fig:YMO_mSR} (b)}.
We calculated ${\bm B}_{\rm loc}$ at Site1$^{(')}$ 
in the incommensurate magnetic structure 
reported in Ref.\ \cite{JHKim2008}.
In this magnetic structure, 
${\bm B}_{\rm loc}$ at Site1$^{(')}$  varies along the $a$- and $c$-axes.
The filled symbols in Fig.~\hyperref[fig:YMO_mSR]{\ref*{fig:YMO_mSR}  (d)} represent the 
calculated values of ${\bm B}_{\rm loc}$ at Site1$^{(')}_{i}$ ($i$ = 1 $\sim$ 20) along the $c$-axis.
The mean and dispersion are considerably smaller than the experimental values, which is 
common to the case of the CM phase.
This result led us to presume that the spin polarization at the O2 sites persists in the LT-ICM phase.
The ${\bm B}_{\rm loc}$ calculated under the assumption that
0.40(1) $\mu_{\rm B}$ and 
0.36(1) $\mu_{\rm B}$ at the O2$_i$ and O2$_{i^{'}}$ sites parallel to 
the local magnetic field at each site is represented by open symbols in Fig.\ \hyperref[fig:YMO_mSR]{\ref*{fig:YMO_mSR}  (d)}; these results are in reasonable agreement with those obtained experimentally.

\begin{table*}[t!]
\centering
\caption{Magnitude and direction of local magnetic field ($B_{\rm loc}$) at Site1$^{(')}_{i}$ ($i$ = 1,2) 
at $T$ = 32 K in the CM phase
of YMn$_2$O$_5$, as obtained from ZF-$\mu$SR experiments and calculations. 
$B^{\rm calc, Mn}$ and $B^{\rm calc, Mn,O}$ represent
the local magnetic field calculated for the Mn and O magnetic structures shown in Fig.\ \hyperref[fig:YMO_mSR]{\ref*{fig:YMO_mSR} (c)} assuming zero and finite O magnetic moments, respectively. 
The direction of $B_{\rm loc}$ at Site1$_2$ and Site1$^{'}_{2}$ is tilted by $\sim 10^{\circ}$ from the $a$-axis.
The magnitude is represented by the
corresponding muon spin precession frequency ($\gamma_\mu B_{\rm loc}/2\pi$).}
{\tabcolsep = 2mm
 \begin{tabular}{c c c c c c c}  \\ \toprule
Site & $ B^{\rm exp}$  (MHz)  & $B^{\rm exp}_{\rm dir}$ &  $B^{\rm calc, Mn}$ (MHz) & $B^{\rm calc, Mn}_{\rm dir}$ & $B^{\rm calc, Mn,O}$ (MHz) & $B^{\rm calc, Mn,O}_{\rm dir}$  \\ \midrule
1$_1$ &  70.3(4) & $c$-axis & 54(1)& $c$-axis & 70(1) & $c$-axis \\
1$_2$ &  70.0(5) & $a$-axis & 39(1) & $\sim a$-axis & 70(2) & $\sim a$-axis \\
1$^{'}_{1}$ &  27.4(5) & $c$-axis & 41(1) & $c$-axis & 27(1) & $c$-axis \\
1$^{'}_{2}$ &  70.3(4) & $a$-axis & 37(1) & $\sim a$-axis & 70(2) & $\sim a$-axis \\ \bottomrule
\end{tabular}}
\label{tb:YMO_calc_site1_Mn}  
 \end{table*} 

Wakimoto {\it et} {\it al}.\ 
reported in a recent polarized neutron scattering study
that the magnitude of vector spin chiral components 
(${\bf S}_i \times {\bf S}_j$) increases in the LT-ICM phase
\cite{SWakimoto}.
Figure \hyperref[fig:YMO_mSR]{\ref*{fig:YMO_mSR}  (e)} 
shows the thermal evolution of 
the difference in neutron cross section between 
scattered spin up ($\sigma_+$) and down ($\sigma_-$) neutrons,
which is in proportion to the component of (${\bf S}_i \times {\bf S}_j$),
indicating that the spin cycloidal structure of Mn$^{4+}$ develops with decreasing temperature
(see Ref.~\cite{SWakimoto} for details).
The temperature dependence of the electric polarization 
is also plotted in Fig.\ \hyperref[fig:YMO_mSR]{\ref*{fig:YMO_mSR}  (e)} for comparison.
For YMn$_2$O$_5$, the electric polarization in the LT-ICM phase 
is assumed to be mainly driven by the cycloidal spin chains of Mn$^{\rm 4+}$ via the SC mechanism,
in which spin vector chirality plays a key role.
Figure \hyperref[fig:YMO_mSR]{\ref*{fig:YMO_mSR}  (e)}  also shows
the thermal evolution of spin polarization at O2 sites, as obtained in the present study.
On the basis of these data, we concluded that the amplitude of spin polarization at O2 sites
increases in the LT-ICM phase with the (${\bf S}_i \times {\bf S}_j$) components.
Xiang {\it et} {\it al}.\  have indicated theoretically 
for the spiral magnet LiCuVO$_4$  
that spin--orbit (SO) coupling on the Cu sites 
drives the asymmetric distribution of electron density mainly around the O atoms \cite{HJXiang2007}.
Thus, given that the spin cycloid chain of Mn$^{4+}$ is formed via O2 ions 
(see Fig.\ \hyperref[fig:YMO_mSR]{\ref*{fig:YMO_mSR}  (c)}), 
our results strongly suggest that 
the SC mechanism is responsible for the imbalance of the charge transfer from the O2 to the Mn$^{4+}$ ions
via the $p$--$d$ hybridization state induced by SO coupling on the Mn sites, thus leading to
the local electric and spin polarization of O2 sites in YMn$_2$O$_5$.

Meanwhile, SmMn$_2$O$_5$ exhibits a perfect collinear magnetic structure in the CM phase, which has no spin vector chiral component, suggesting the absence of charge transfer between O2 and Mn$^{4+}$ ions in this compound.
In fact, no evidence was found in the present study for spin polarization of O2 ions in the CM phase.
Thus, we concluded that ionic displacement via the ES model is the main origin of the ferroelectricity in SmMn$_2$O$_5$.

\section{\label{sec:level4} Summary}
We successfully observed
spin polarization of O2 ions in the multiferroic phase of YMn$_2$O$_5$ via the synergetic use of RSXS and $\mu$SR techniques.
The amplitude of O spin polarization shows a remarkable increase in the LT-ICM phase in proportion to the vector spin chiral components, implying that the "spin current" model is the most likely scenario for the electronic displacements that result in ferroelectricity with a spin cycloidal structure in multiferroic materials.
The present study thus leads to further understanding of electromagnetic coupling in multiferroic materials 
and advances the development of multiferroic device applications. 
\section{\label{sec:level5} Acknowledgement}
We would like to express our thanks to the TRIUMF staff for their technical support during the $\mu$SR experiments and to H. Lee for his assistance with the DFT calculations. 
This study was supported by the KAKENHI program for Scientific Research (A) (JP15H02038, JP17K05130) and (B) (24340064), Challenging Exploratory Research (2365409), Dynamic Alliance for Open Innovation Bridging Human, Environment and Materials, and Condensed Matter Research Center, IMSS, KEK.
This work was performed with the approval of the Photon Factory Program Advisory Committee 
(Proposals No. 2017G549, No. 2017PF-BL-19B,  and No. 2019PF-22).


\begin{thebibliography}{36}
\bibitem{Fitting} We analyzed the time spectrum data using the following equations: $A(t) = \sum_{i=1}^{m} A_i \cos(\omega_i t+\phi)\exp(-\lambda_it)+C $ for the CM phase and  $A(t) = \sum_{i=1}^{m} A_i J_0(\omega_it)\exp(-\lambda_it)+C $ for the ICM phase, where $A_i$ is the partial asymmetry, $\omega_i$ is the muon spin precession frequency, $\phi$ is the initial phase of precession, $\lambda_i$ is the transverse relaxation rate, $C$ is a constant component, and $J_0 (x)$ is the Bessel function.
\end{thebibliography}
\end{document}